\documentclass{article}
\usepackage{ijcai20}

\usepackage{times}
\usepackage{soul}
\usepackage{url}
\usepackage[hidelinks]{hyperref}
\usepackage[utf8]{inputenc}
\usepackage[small]{caption}
\usepackage{graphicx}
\usepackage{amsmath}
\usepackage{amsthm}
\usepackage{booktabs}
\usepackage{algorithm}
\usepackage{algorithmic}
\usepackage{xcolor}
\usepackage{balance}

\title{An Information Diffusion Approach to Rumor Propagation and Identification on Twitter}

\author{
Abiola Osho$^1$
\and
Caden Waters$^1$\And
George Amariucai$^1$
\affiliations
$^1$Kansas State University\\
\emails
aaarise@ksu.edu,
crwaters@ksu.edu,
amariucai@ksu.edu
}

\begin{document}

\maketitle

\begin{abstract}
With the increasing use of online social networks as a source of news and information, the propensity for a rumor to disseminate widely and quickly poses a great concern, especially in disaster situations where users do not have enough time to fact-check posts before making the informed decision to react to a post that appears to be credible. In this study, we explore the propagation pattern of rumors on Twitter by exploring the dynamics of microscopic-level misinformation spread, based on the latent message and user interaction attributes. We perform supervised learning for feature selection and prediction. Experimental results with real-world data sets give the models' prediction accuracy at about 90\% for the diffusion of both \textit{True} and \textit{False} topics. Our findings confirm that rumor cascades run deeper and that rumor masked as news, and messages that incite fear, will diffuse faster than other messages. We show that the models for \textit{True} and \textit{False} message propagation differ significantly, both in the prediction parameters and in the message features that govern the diffusion. Finally, we show that the diffusion pattern is an important metric in identifying the credibility of a tweet.

\end{abstract}

\section{Introduction}
Online social networks (OSNs) like Twitter have become  increasingly popular for dissemination of information, news, and events around the world. Due to their wide reach, oversimplified conversations, and ability to provide quick blasts of information, online social networks have also become an avenue for the spread of rumors. With the current political and economic climate around the world, we continue to witness the spread of falsehood, and pandemic strikes initiated in 280 characters or less. In the absence of verification sources, individuals can use online media to disperse and coordinate information, since the potential spread of information (whether true or false) is impartial to the content or source.

The impartial and unrestrained spread of information in social networks can be of great value as observed in September 2015 where the US geological survey tracked earthquakes by simply following mentions of the term 'earthquake'\cite{usgs}, or the 2012-13 flu epidemic where researchers used tweet data to correlate the spread of the disease with a view to reducing its impact \cite{broniatowski2013national}, and in stock markets where consumer insights companies use social media data to predict shifts in consumer spending behaviors that translate to shifts in stock price. However, it becomes detrimental when the information is false, like during hurricane Sandy where there were false tweets about the NYSE being flooded with up to 3 feet of water, which even got reported by some news outlets \cite{sandy}.

According to deflationism \cite{soames1997truth}, assertions that predicate truth of a statement do not attribute a truth property to such a statement . Since there is no real-world truth label to posts (i.e., text, images, memes, etc.), OSN users simple decide to react to a post based on the perceived credibility of the message. A message intended to deceive might have concealed meanings, emotions and sentiments even if it does not come off as one. The search for truthfulness of a message might be lacking depending on how accepting or prejudiced the user is towards a topic, especially when they are exposed to contradicting information from diverse sources. Since some rumors never completely die out, persisting with low frequencies with potential for flare-ups from time to time, we hypothesize that there is a difference in the spreading behavior of rumor and truthful information in OSNs, a difference dependent on the latent attributes of the message, along with the interaction between users encountering these messages. Seeing as feature design and selection strongly impact a machine learning model's accuracy much more than the model used \cite{hall1999correlation}, we train a Bayesian Logistic Regression model by incorporating network, interaction and message features to measure the node-to-node influence dynamics to rumor propagation. We extract observable and latent features from the Twitter API as independent variables to the predictive model and present a ranking of the features significantly impacting diffusion of rumors.


Existing research in rumor propagation and identification examine the behavior of misinformation posts over the network based diffusion speed, depth, concentration, location, and sometimes combining features to differentiate posts. However, with access restrictions to the complete Twitter network graph and posts, it is important that we examine how individual users contribute to the diffusion on rumor posts and what features of the post sharer and receivers influence this paradigm. Since the spread of gossip is a uniform process, spreading from node to node \cite{pittel1987spreading}, it is essential to note that the diffusion process is influenced not only by the creator of the tweet, but also by the sharer of the tweet. In this paper, we investigate the problem of rumor propagation by exploring the dynamics of microscopic-level misinformation spread, based on the latent message and user interaction attributes for the consumption and sharing of misinformation in online social networks. By observing the spreading behavior of rumors in online social networks, we propose a model that predicts a user's reaction to a rumor, based on the hidden meanings of the message, combined with their interaction with the spreading user. The approach is to (1) identify topics labeled as \textit{False} (in other words, rumor) or \textit{True} using \textit{Snopes} \cite{Snopes} -- an online fact-checking site -- and collect Twitter posts about the topic. (2) To each user, we associate a total of 17 features, to include 3 network and 14 interaction attributes; and to each message, we associate 24 attributes: 10 observable and 14 latent attributes. (3) For each pair of users with established followership relationship, we assign a diffusion label: \textit{diffused}, if the follower has reacted to the message or \textit{not diffused}, if the follower ignores the message (4) We train a random forest classifier to rank top features that differentiate the propagation pattern between rumor and normal posts. (5) We train a Bayesian logistic regression model with the top-20 ranked features to predict whether a user will react to a rumor based on his interaction with the friend sharing the post. We further extend the model to predict the truth-status of a tweet, based on the propagation pattern of the message. By adopting the same set of features, we include the diffusion tag as an input variable to the training model. The contributions of this paper are as follows:
\begin{itemize}
    \item We introduce 8 new latent message attributes for misinformation propagation and identification.
    \item Using these new attributes, we present a microscopic-level rumor diffusion model based on network, interaction and message features of users encountering the posts, and achieve an accuracy up to 10\% higher that previous models.
    \item We present a model to predict the truth-status of a tweet using the propagation pattern, and achieve an accuracy up to 10\% higher than models that do not incorporate propagation patterns.
    \item We develop a pre-trained model which can be used `as-is' for predicting diffusion of misinformation on Twitter.
\end{itemize}



The paper is organized as follows: Section \ref{prevwork} reviews the related work on misinformation diffusion, and features that aid misinformation spread in social networks. Section \ref{method} describes our general approach, feature selection, and classification algorithms. Section \ref{exp} elaborates on the experiment, data used, prediction model and evaluation metrics. Section \ref{result} presents experimental results and observations, and finally, Section \ref{summary} gives conclusions and insights into possible future works.

\section{Related Work}\label{prevwork}
The spread, detection and control of true and false news online continues to be a topic of interests to researchers in humanities, social sciences and engineering. In this section, we give insight to some of the studies and methods relating to rumor propagation and detection.

\subsection{Information Diffusion}
The information diffusion process can be observed through the diffusion graph and rate of adoption of the information by the nodes in the graph. The diffusion graph shows influence in the network, which is important for viral marketing \cite{subramani2003knowledge} \cite{chen2010scalable}\cite{domingos2005mining}, crisis communication \cite{acar2011twitter} and  retweetability\cite{neppalli2016retweetability}. Generally, influence analysis models have focused on relationship strength based on profile similarity and interaction activity \cite{xiang2010modeling}, and the mechanisms responsible for network homogeneity \cite{lewis2012social}. Identifying influential users has been found to be useful when trying to select seed nodes in the community that will maximize the spread of information across the networks. \cite{3} worked on finding the best spreaders in dissimilar social platforms when the complete global network structure is unavailable. The work of \cite{11} observed that (1) the authority of an  influential user on social media which can be used to change the opinions of other users and (2) opinion similarity factors where users tend to accept an opinion that is similar to his own, are important factors when selecting seed nodes for information spread.

\subsection{Rumor Propagation}
Research in political science explored the differential diffusion of true, false, and mixed (partially true, partially false) news stories on Twitter using the fact-checked rumor cascades that spread on Twitter over a 12-year period. \cite{vosoughi2018spread} observed that falsehood diffused faster, farther, deeper and more broadly than truth in all categories of information, with a more noticeable impact in false political news. The study also observed that false news are often more novel, inspiring fear, disgust and surprise in replies while true stories inspired anticipation, sadness, joy and trust. In like manner, \cite{grinberg2019fake} examined the spread of fake news on Twitter during the 2016 U.S. presidential election and observed that the exposure to fake news sources was extremely concentrated with seven fake news sources accounting for more than 50\% of fake news exposures. The study showed that political affinity was associated with the sharing of content from fake news sources and that the sharing of content from fake news sources was positively associated with tweeting about politics, and exposure to fake news sources. Computer scientists like \cite{friggeri2014rumor} examined the spread of rumors on Facebook and found that rumor cascades run deeper in the social networks. When rumor debunking posts are available, \cite{takahashi2012rumor}\cite{friggeri2014rumor} reported that users will either delete a post, if it is confirmed to be rumor, or share otherwise. \cite{abdullah2015user} revealed that users spread the messages which they think is important and mostly retweet messages because of the need to retweet interesting tweet content or tweet creator.

\subsection{Feature-based rumor detection}
\cite{castillo2011information} extracted 68 features from tweets and categorized them as (1) message-based which considers characteristics of the tweet content, such as length of post, presence of exclamation, number of positive/negative sentiment words, (2) user-based which considers characteristics of Twitter users, such as registration age, number of followers, number of friends, and number of user posted tweets, (3) topic-based which aggregates the message-based and user-based features, and (4)
propagation-based which considers characteristics related the propagation tree that can be built from the retweet of the post. \cite{liang2015rumor} explored rumor identification using users' behavior to differentiate between normal authors and rumormongers. \cite{wu2015false} introduced the propagation tree, and used a random walk graph-kernel based hybrid SVM classifier to capture the high-order propagation patterns in addition to topic and sentiment features for rumor detection in Sina Weibo.  \cite{yang2012automatic} proposed two new features (1) client-based feature referring to mode of access, whether mobile or non-mobile, and (2) location-based feature referring to the actual place where the event mentioned by the rumor-related microblogs happened, domestic (in China) and foreign. \cite{kwon2013prominent} observed from rumor time series that rumors tend to have multiple and periodic spikes, whereas non-rumors typically have a single prominent spike, proposed an automatic detection of rumor on Social Network using Periodic External Shocks model to detect rumors. \cite{mendoza2010twitter} analyzed the retweet network topology and found the diffusion patterns of rumors different from news. They also found that rumors tend to be questioned more than news by the Twitter community, suggesting that the Twitter community works as a collaborative filter of information.

In our work, we focus on examining how rumor is propagated and the hidden qualities of the message useful in identifying messages laced with falsehood, and how the propagation style can help in identifying rumor.

\section{Features for rumor propagation and identification}\label{method}
Here, we describe a framework, where given tweet, will predict (1) whether the followers of the spreader (could be the author or someone sharing) will react to the tweet in the form of a retweet, share, quote, like or favorite, and also (2) use the predicted propagation pattern to determine the truth-status (if the post is a rumor or not). We suggest 3 categories of features: message, interaction, network, and train a random forest classifier to rank the features in order of importance, then we build a Bayesian logistic regression model for classification. We adopt some of the features examined in literature and suggest new ones, described below.

\subsection{Network-based features}
In microblogs such as Twitter, a friend is someone a user follows, and a user can see all of his friend's post, in like manner, a follower is someone that follows and has direct access to all of a user's post. We consider three features of the user's network: \textit{followers count}, \textit{friend's count}, which have been extensively studied by \cite{castillo2011information}\cite{liang2015rumor}\cite{yang2012automatic}, and \textit{followers to friends ratio}, which was used in \cite{wu2015false} to establish opinion leaders. These attributes are important because a user's friends impact the kind and volume of messages that end up in his timeline and the higher the number of followers, the farther the possibility of reach. This is also reflected in policies by OSNs like Twitter and Instagram who attach value to the followers count, where users become verified once they cross a certain threshold, even if the account holder is not a celebrity or public figure. Table \ref{network} describes the network features used in the model.

\begin{table}[ht]
\footnotesize
\centering
\begin{tabular}{p{1in}p{2in}}   
\toprule
Feature  &  Description \\
\midrule
followers count & higher count depict higher reach\\
friends count & \# of accounts user follows\\
followers-friend & ratio to show influence in the network \\
\bottomrule
\end{tabular}
\caption{Network-based features}
\vspace{-3mm}
\label{network}
\end{table}

\subsection{Interaction-based Features}
Since we are exploring rumor propagation as being dependent on the influence being wielded between users and taking propagation depth to be a factor of how messages cascade across the network, we examine the nitty-gritty of the followee-follower relationship to establish the features that influence the spread of rumor over the network. Here, we identify specific attributes of the user's online persona and posting behavior as determinant to being an influencer or influenced in the network. The assumption is that both the follower and followee contribute equally to the diffusion of a post, and an aggregate of network and message attributes tilt the reaction decision. Table \ref{interaction} describes the 14 interaction attributes being considered. The last 5 features have been explored by \cite{castillo2011information}, while we introduce 9 new features to the study of rumors in social networks.

\begin{table}[ht]
\footnotesize
\centering
\begin{tabular}{p{0.9in}p{2.1in}}
\toprule
Feature  &  Description \\
\midrule
shared friends & common nodes they interact with\\
directed tweets & ratio of tweets directed at someone\\
dialogue & active interaction from user 1 to 2\\
retweet-to-tweet & ratio of user's tweets with retweet\\
tweet wit hashtag & ratio of user's tweets that contain hashtags\\
tweets with url & ratio of user's posts with URL\\
tweets with media & ratio of user's posts with media\\
avg favorite-tweet & ratio of posts that get favorited\\
avg tweets/day & shows how active the user is\\
has url	& does user's profile have a URL\\
has description & does user's profile have description\\
is verified	& is the account verified\\
status count &	volume of tweets over account's lifetime\\
account age	& \# of days since account was created\\
\bottomrule
\end{tabular}
\caption{Interaction-based features}
\vspace{-3mm}
\label{interaction}
\end{table}

\subsection{Message-based Features}
Twitter posts are very fluid, taking up varying forms as feedback, news, marketing campaigns, etc., so it is expected that rumor in this medium come in all forms. We account for this variation and consider the concealed form and intents of posts. Previous work have focused on count of positive and negative words in a tweet, with some exploring the polarity of the message sentiment but we look to explore the latent attributes of the message by introducing new features encompassing the type of post and emotion it is meant to incite. We adopt paralleldots API to perform content analysis on tweets to reveal the sentiment, intent, emotion and abusive attributes. To the best of our knowledge, this is the first time emotion and intent will be introduced as attributes of the message for identifying rumors in microblogs. Table \ref{message} describe the message attributes adopted in our model. We introduce the last 12 features as latent features relating to the form, meaning and intent of the message.

\begin{table}[ht]
\footnotesize
\centering
\begin{tabular}{p{1in}p{2.1in}} 
\toprule
Feature  &  Description \\
\midrule
quoted status & has post been quoted \\
is rt& has post been retweet\\
rt count& \# of retweets\\	
rt status & is post a retweet \\
favorited count& \# of favorites\\
has hashtag& does post contain hashtags\\
has url	& does post contain URL\\
has mentions& does post mention someone using ``@"\\
has media& does post contain media\\
avg tweet length& length of tweet / 280 (max length)\\
positive sentiment& positive polarity of tweet\\
negative sentiment& negative polarity of tweet\\
neutral sentiment& neutral polarity of tweet\\
happy emotion& is post meant to incite happiness\\
fear emotion& is post meant to incite fear\\
sad emotion& is post meant to incite sadness\\
angry emotion& is post meant to incite anger\\
bored emotion& is post meant to incite boredom\\
feedback intent& is post meant to be a feedback\\
news intent& is post meant to be news\\
query intent& is post meant to be a query\\
spam intent & is post meant to be spam\\
marketing intent& is post meant for marketing\\
abusive& is post abusive\\
\bottomrule
\end{tabular}
\caption{Message-based Features}
\vspace{-3mm}
\label{message}
\end{table}

\section{Experiment}\label{exp}
In this section, we describe the data collection process, prediction models and the metrics for evaluation.

\begin{table*}[ht!]
\footnotesize
\centering
\begin{tabular}{p{0.35in}p{3.0in}p{2.3in}p{0.44in}} 
\toprule
Category & Topic & Keywords & \# tweets\\
\midrule
False   & Hillary Clinton said ``we must destroy Syria for Israel" & Hillary, destroy, syria & 96724\\
        & The FBI discovered bones of young children in Jeffrey Epstein's private island & epstein, bones, children & 189812\\
        & Odessa shooter was ``a Democrat Socialist who had a Beto sticker on his truck." & odessa, shooter, beto, democrat, sticker & 19491\\ 
\midrule
True    & Blood spots visible in the left eye of Joe Biden during a CNN debate in Sept. 2019 & Joe, Biden, blood, eye &  38983\\
        & Anti-abortion Rep. DesJarlais encouraged some women to have abortions & abortion, Desjarlais, mistress, republican & 15591\\
        & Video shows air traffic over the US on 9/11 as thousands of flights were grounded after a terrorist attack & flights, grounded, after, 9/11 & 10355\\
\bottomrule
\end{tabular}
\scriptsize\caption{Topics identified from Snopes, along with the associating keywords used in querying the Twitter seach API}
\vspace{-5mm}
\label{tab:topics}
\end{table*}

\subsection{Data Collection}
We used Snopes \cite{Snopes} to identify topics that have been fact-checked and rated as \textit{True} or \textit{False}. Even though Snopes has different categories including those labeled ``Mostly True" and ``Mostly False", we restrict this research to those that are strictly labelled \textit{True} or \textit{False}. 

For each topic, we assign a set of keywords and crawl the Twitter search API using queries of the form ($K_1 \vee K_2 \vee K_3$), similar to that described by \cite{mathioudakis2010twittermonitor} but with $K_i$ representing the conjunction of possible keyword combinations. For instance, the topic ``In a leaked e-mail, Hillary Clinton said ``we must destroy Syria for Israel."" had keywords ``hillary, destroy, syria" and query $((hillary \wedge destroy \wedge syria) \vee (hillary \wedge destroy) \vee (hillary\wedge syria) \vee (destroy \wedge syria))$. Table \ref{tab:topics} gives a breakdown of our topics, along with the associating keywords and number of tweets (including retweets). For reproducibility and future adoption, we make the crawler publicly available for researchers on GitHub.

In the dataset, we found a large variation in the volume of tweets in the \textit{True} and \textit{False} collections, with \textit{False} posts accounting for more than 80\% of the entire dataset. Also, we observed that the propagation depth of \textit{False} posts ran deeper with an average of 4 retweet depth while \textit{True} posts averaged 2 retweet depth. Lastly, we observed a ``diffused"/``not diffused" ratio of 35/65 for the tweets in the collection of \textit{True} topics and 45/65 for tweets in the collection of \textit{False} topics. 


\subsection{Prediction Models}
Given a collection of messages and the associated user, we recreate the Twitter followership graph by connecting all of the user's followers. Based on the assumption that users will interact with their friends' messages uniquely, we assign the diffusion label as a function of the reaction observed per message. To show that the microscopic-level misinformation spread based on the latent message and user interaction attributes is sufficient to give insight to the credibility of a message, we perform two supervised learning tasks by adopting two off-the shelf machine learning models: Bayesian Logistic Regression and Random Forests for prediction and feature selection, respectively. 

\noindent
\textbf{Predicting rumor propagation:} We perform a node-to-node analysis between a pair of users, the spreader and receiver, examining each user's posting behavior, and their interactions to predict the receiver's reaction. First, we built a Random Forests classifier to analyse the importance of the input features and perform selection on the best features for rumor propagation and identification tasks. Then, we built separate models for \textit{True} and \textit{False}, performed supervised learning task using the Bayesian logistic regression by assigning diffusion label ``diffused" between a spreader and his follower, if the follower has reacted to an identified tweet (in either case, \textit{True} or \textit{False}) and ``not-diffused" otherwise. We adopt an 80-20 train-test split of the data and account for over-fitting by performing 10-fold cross validation. We make predictions on the capability of the model to correctly predict diffusion on the message type and take it a step further by investigating the model's ability to generalize across message type. 

\noindent
\textbf{Predicting credibility of posts:} Here, we perform operations similar to predicting rumor propagation, but train a model that predicts if a message is \textit{True} or \textit{False}, by including the diffusion property as an independent variable during the training phase. We examine how users on Twitter relate with posts of their friends by building classifiers to distinguish user interactions based on the credibility of the message. For a message $m$, where $m \in \{1,\ldots,M\}$, spread over a network with $n$ interactions, we train a model that predicts the truth status of the message based on the diffusion behavior observed along each one of the $n$ links along which the message propagates. The predicted output is the majority truth status observed across the $n$ interactions. For instance, if a \textit{True} message is spread over 5 interactions and the model predicts the post to be \textit{True} 3 out of 5 times, we accept the output to be \textit{True} and evaluate the model over its correct classification of $M$ messages in the test collection.

\subsection{Evaluation Metrics}
The prediction capabilities of the learned model are tested based on its abilities to predict if there is diffusion across an edge given the learned model. We use standard classification evaluation metrics: precision, recall and F score, to assess the efficiency of our model.

\textbf{Precision} describes the ratio of instances correctly classified as ``diffused'' to the total classified as ``diffused'', and is estimated as:

\begin{equation}
     Precision = \frac{TP}{TP + FP}
\end{equation}

\textbf{Recall} is the ratio of instances correctly classified as ``diffused'' to the total number of instances that ``diffused", and is estimated as:
\begin{equation}
     Recall = \frac{TP}{TP + FN},
\end{equation}

\noindent
Where $TP$ (true positives) is the number of instances correctly classified as ``diffused", $FP$ (false positives) is the number of instances incorrectly classified as ``diffused", and $FN$ (false negatives) is the number of the instances incorrectly classified as `` not diffused".

\vspace{1.5mm}
\textbf{F score} is the harmonic mean of the precision and recall. It is computed as 
\begin{equation}
      F score = 2 \times \frac{Precision \times Recall }{Precision + Recall}
\end{equation}

\section{Results}\label{result}
In this section, we report the results obtained from each phase of the experiment.

\subsection{Features analysis for rumor propagation}
One justification for using multivariate methods is that they take into account feature redundancy and yield more compact subsets of features, as features that are individually irrelevant may become relevant when used in combination, which also shows that correlation between sets of features does not necessarily imply redundancy. Considering that the goal of the feature analysis task of this study is to identify the optimal set of features necessary to maximize prediction of misinformation diffusion, we train a random forests model and then select the top 20 features for the learning and prediction tasks. These features are ranked in  Table~\ref{tab:features} in descending order of importance.

\begin{table}[ht!]
\footnotesize
\centering
\begin{tabular}{|c|l|l|} 
\hline
Rank & False  & True \\
\hline
1 & MSG is RT & MSG is RT\\
2 & MSG favorited count & social homogeneity\\
3 & MSG has mentions & MSG favorited count\\              
4 & dest tweet with hashtag & src tweets with URL\\
5 & src retweet-to-tweet & MSG feedback intent\\    
6 & MSG news intent & MSG positive sentiment\\         
7 & src followers count & src directed tweet\\
8 & MSG has URL & MSG has URL\\       
9 & src followers-friends & src avg favorite-tweet\\
10 & src account age & src avg tweet/day\\         
11 & src tweets with URL & src followers count\\              
12 & MSG fear emotion & MSG has mentions \\            
13 & dest tweet with hashtag & src account age\\                
14 & src status count & dest retweet-to-tweet\\                 
15 & src friends count & src retweet-to-tweet\\
16 & social homogeneity & src has URL\\
17 & MSG RT count & dest follower-friends\\
18 & dest friends count & src status count\\
19 & MSG positive sentiment &  MSG has hashtag\\
20 &  MSG negative sentiment & MSG RT status\\
\hline
\end{tabular}
\scriptsize\caption{Top 20 features for predicting misinformation diffusion selected using Random Forest classifiers}
\vspace{-3mm}
\label{tab:features}
\end{table}

Given the attributes describing a user's network and interaction, along with those of the message, we observed that for message with \textit{False} status, message attributes account for 45\% of the ranked features with the combination of network and interaction accounting for 55\%, while message attributes account for 40\% of top ranked features for \textit{True} posts. 

As anticipated, the latent attributes of the message rank in the top features for both \textit{True} and \textit{False} models, confirming that the meaning, intention and emotions of messages influence users' decisions in the diffusion process. From the ranked features, we can infer that rumor posts masked as news, meant to incite fear will diffuse better than others. However, it is surprising that the diffusion of rumor posts cannot be strictly tied to their sentiment as we observed that both negative and positive sentiments contribute equally to the performance of the model. Even though it ranks differently in both models, social homogeneity ranking well in both models shows that a user will most likely respond to the post of someone with interests similar to his own.

\subsection{Predicting Rumor Propagation}
We focus on the problem of predicting the diffusion decision (to react or not) of a user based on his perception of the message and interaction with the spreader of the information. In this model, we do not take into account the effect of previous exposure to similar posts, or the popularity of the message, we simply make an inference on whether a user will retweet, share, quote or favorite a tweet by estimating the probability of diffusion.

In Table~\ref{tab:eval}, we show the performance of the model across message type, using the performance metrics previously highlighted. Using the F as measure of accuracy, we achieved 91.6\% and 89.9\% prediction accuracy for message with \textit{True} and \textit{False} status respectively. 

\begin{table}[ht!]
\centering
\begin{tabular}{p{1.2in}p{0.5in}p{0.5in}p{0.5in}} 
\hline
 Model &  Precision & Recall &  F  \\
\hline
False &  0.897  &  0.902  & 0.899 \\
True &  0.908  & 0.925  &  0.916\\
Credibility & 0.919 & 0.903 & 0.911\\
\hline
\end{tabular}
\scriptsize\caption{Model performance for predicting diffusion of \textit{True} and \textit{False} posts, and credibility status of a message}
\vspace{-5mm}
\label{tab:eval}
\end{table}

\bigskip
\noindent
\textbf{Pre-trained model for \textit{True}-\textit{False} diffusion prediction}
Given current events highlighting the spread of falsehood on social networks, and the increasing importance of transfer learning \cite{transferlearning}, one of the motivations for this research was to develop a model robust enough that it can be adapted as an off-the-shelf prediction model for rumor diffusion in social networks. To do this, we extensively tested our model's performance over topics outside the training list. The results for inter-topic and inter-credibility prediction tasks are reported in Table \ref{tab:inter}. For inter-topic test, we observed performance of similar magnitude in diffusion prediction capabilities when the models are exposed to topics outside the training list. As observed from  the table, there is a difference for inter-credibility test and we believe this is due to the difference in the features that influence diffusion for the message types. However, this difference is not significant enough for us to discard the abilities of these models.

\begin{table}[ht!]
\centering
\begin{tabular}{p{1.2in}p{0.5in}p{0.5in}p{0.5in}}  
\hline
 Model &  Precision & Recall &  F  \\
\hline
False &  0.887  &  0.889  & 0.882 \\
True &  0.899  & 0.919  &  0.908 \\
False model-True test & 0.856 & 0.821 & 0.838 \\
True model-False test & 0.849 & 0.921 & 0.884 \\
\hline
\end{tabular}
\scriptsize\caption{Model performance for inter-topic, inter-credibility diffusion prediction}
\vspace{-5mm}
\label{tab:inter}
\end{table}

\subsection{Predicting credibility by diffusion pattern}
While some users react to posts of varying credibility, others only react to tweets that are precisely \textit{True} or \textit{False}. So training a model that learns to distinguish this interaction-reaction relationship is useful for identifying the credibility of a tweet by observing the reaction of a user based on the established interaction between the users. By incorporating the diffusion status of a tweet, we train a  model to predict the credibility of the message. The objective of the task is to show that the diffusion behavior influenced by the interaction between users is useful for predicting the credibility of a post.
The result from our experiment validate our assumption that the difference associated with the message, interaction and diffusion patterns of \textit{True} and \textit{False} posts can be exploited in predicting the credibility of messages. By combining these attributes, we were able to achieve 91\% accuracy in identifying whether messages are credible or not, see Table \ref{tab:eval}. It is important to note that the model is tested using labelled data with existing ground-truth. To show the impact of the diffusion attribute to the credibility prediction task, we carried out a parallel credibility identification task without the diffusion label and observed a performance of 82\%. 


\section{Summary}\label{summary}
In this study, we hypothesized that the diffusion of rumor on Twitter is significantly influenced by the latent attributes of the message, coupled with the interaction between users encountering the message. We exploited Twitter relationship as the foundation for the diffusion of misinformation and showed that the hidden meaning of a message plays an important role in the reaction decision of a user. Also, we showed that rumor is mostly masked as news content, meant to incite fear emotions in the reader with mixed sentiments, and that the diffusion attribute is significant to predicting the credibility of a tweet. In the future, we hope to adapt this model to topics that have mixed content with hopes to identify how much of the message is true.

\balance
\bibliographystyle{named}
\bibliography{misinform}

\end{document}